# Recursive Neural Language Architecture for Tag Prediction

**Saurabh Kataria**
Palo Alto Research Center
Webster, NY - 14580

## Abstract

We consider the problem of learning distributed representations for tags from their associated content for the task of tag recommendation. Considering tagging information is usually very sparse, effective learning from content and tag association is very crucial and challenging task. Recently, various neural representation learning models such as WSABIE and its variants show promising performance, mainly due to compact feature representations learned in a semantic space. However, their capacity is limited by a linear compositional approach for representing tags as sum of equal parts and hurt their performance. In this work, we propose a *neural feedback relevance model* for learning tag representations with weighted feature representations. Our experiments on two widely used datasets show significant improvement for quality of recommendations over various baselines.

## Introduction

With the advent of Web 2.0 and explosive growth of user generated content, tagging –a social bookmark indexing activity– has become a preferred medium for conceptually organizing and summarizing information. However, from a user perspective, it is often difficult to compose a set of words that can represent associated content. Moreover, from a content sharing website perspective, tagging information can quickly become inconsistent and idiosyncratic because of disparate phrasing styles of the users, which leads to ineffective utilization of tag information. Therefore, automatic tag recommendation has become a popular choice both for content sharing websites as well as its users.

Tag recommendation remains a challenging task, mainly due to extremely sparse tag information present for the underlying content. Recent advances in learning latent representations, or *embeddings* based upon various neural architectures are proving to be successful in tag recommendation tasks [Wang *et al.*, 2015; Weston *et al.*, 2014], mainly due to compact representations of words and tags in continuous low dimensional space. Unsupervised word embedding methods train with a reconstruction objective in which the embeddings are used to predict the original text. For example, word2vec [Mikolov *et al.*, 2013] tries to predict all the words in a document, given the embeddings of surrounding words. As a result, word embeddings carry semantic information where words that are contextually similar are "closer" compared to dissimilar ones. In contrast, supervised embedding methods, e.g. WSABIE [Weston *et al.*, 2011] and its variants [Weston *et al.*, 2013b], embed both the labels (or tags) and documents in a shared semantic space where document embedding predicts the "closest" label embedding. The document embeddings are obtained by combining the embeddings of its words using a model-dependent, possibly learned function, producing a point in the same embedding space.

Although supervised embedding methods are easy to implement and comprehend, these models are limited by a simplistic assumption that each underlying word embedding contribute equally to the document embedding. Consequently, the burden lies on the ranking objective (e.g. WARP [Weston *et al.*, 2011]) to discriminate among the relevant and non-relevant features while learning and making a prediction, causing the model to underfit. Moreover, in the case of tag recommendation, ambiguity among certain tags (e.g. "apple" as a tag in context of a technology article vs. a nutrition article) makes the learning even harder as two separate classes of words compete to achieve same representation for an ambiguous tag.

In this work, we relax the simplistic compositional assumption where word embeddings combine linearly to form tag embeddings. Specifically, we extend supervised linear embedding architecture (e.g. WSABIE) with a *neural relevance feedback layer* that weights each word in a document based upon its relevance to the tag it is composed to. Our formulation is flexible as it can accommodate bilinear network layer that can implicitly assign types to words and tags representations based on different contextual usage and calculate their relevance multiplicatively. Moreover, parameters of the relevance layer can be updated jointly with supervised tag embedding layer providing an easy and scalable learning approach.

Contributions of our work are as follows: we propose a relevance feedback based extension to popular supervised representation learning framework a.k.a WSABIE. Our proposed approach works by reweighting, using the relevance function, each part of the sum (i.e., words in a document) that comprises the embedding of tags for a given set of documents. We propose several classes of neural networks based relevance functions that capture similarity between a word in document and its associated tag. As a result, we show that relevance functions with higher capacity can help disambiguating context in which a tag is associated with a document and, therefore, helps improving tag recommendation tasks. Lastly, we apply our relevance feedback based neural network to tag recommendation tasks to two publically available data set and show significant improvements over various baselines.

## Related Work

We divide related work in two areas of research:(1) distributed representation learning from content and tags, and (2) tag recommendation for documents.

**Distributed Representation Learning from Content and Tags** Earlier research in distributed representation learning [Bengio *et al.*, 2003] has focused on using probabilistic neural networks to build general representations of words that improve upon the classic n-gram language models. More recently, this approach has been extended with two popular neural language models [Mikolov *et al.*, 2013] for learning distributed representations of words, that are (1) continuous bag-of-words model (CBOW) and (2) Skip-gram, collectively known as *word2vec*. Although Word2vec and its document based extensions [Le and Mikolov, 2014], learns unsupervised embeddings that are shown to be successful for various NLP related tasks [Djuric *et al.*, 2015; Le and Mikolov, 2014], supervised embeddings such as WSABIE [Weston *et al.*, 2011] directly learns from tag-word associations and outperform word2vec for prediction tasks [Weston *et al.*, 2014].WSABIE has been shown to perform well on various recommendation tasks as well, e.g., music annotation with textual tags [Weston *et al.*, 2012], personalized video recommendation [Weston *et al.*, 2013a], image annotation with labels, i.e. ImageNet [Weston *et al.*, 2013c], personalized tag recommendation for images [Denton *et al.*, 2015]. [Weston *et al.*, 2014] extended WSABIE with a convolutional neural network based document representation that can take word ordering into account in a supervised embedding framework.

**Tag Recommendation for Documents** Tag recommendation methods can roughly be categorized into three classes [Wang *et al.*, 2012]: content-based methods, co-occurrence based methods, and hybrid methods. Content-based methods [Shen and Fan, 2010; Landia, 2012; Lu *et al.*, 2009] utilize only the content information (e.g., abstracts of articles, image pixels, and music content) for tag recommendation. Co-occurrence based methods [Garg and Weber, 2008; Rendle *et al.*, 2009] are similar to collaborative filtering (CF) methods [Li and Yeung, 2009]. The co-occurrence of tags among items, usually represented as a tag-item matrix, is used for tagging. The third class of methods [Wu *et al.*, 2009; Wang and Blei, 2011; Wang *et al.*, 2013; 2015], also the most popular and effective ones, consists of hybrid methods. They make use of both tagging (co-occurrence) information (the tag-item matrix) and item content information for recommendation.

Learning item representations becomes crucial in tag recommendation especially when the tag-item matrix is extremely sparse. Recently, models such as collaborative topic regression (CTR) [Wang and Blei, 2011] and its variants [Wang *et al.*, 2013; Purushotham *et al.*, 2012] have been proposed and adapted for tag recommendation to achieve promising performance. These models use latent Dirichlet allocation (LDA) [Blei *et al.*, 2003] as the key component for learning item representations and use probabilistic matrix factorization (PMF) [Salakhutdinov and Mnih, 2008] to process the co-occurrence matrix (tag-item matrix). Although powerful approaches for tag recommendation, CTR and its variants suffer from inconsistent low dimensions corresponding to content topic space and co-occurrence matrix's latent factor space which typically have different sparsity and capacities. Deep learning based generative models, such as deep generative autoencoders [Wang *et al.*, 2015] remedy this drawback by representing both content and tag co-occurrences in same lower dimensional space and show improvements over CTR. However, a cubic learning time complexity (in terms of low dimensional space) for these models prohibit effective posterior estimation and resort to approximations [Li and Yeung, 2009; Wang *et al.*, 2015].

## Approach

In this section we present our approach for learning distributed representations for tags and words in a common, low-dimensional embedding space. We consider the learning setting where training documents are annotated with their tags. Formally, we are given a corpus $\mathcal{M}$ of $|\mathcal{M}|$ documents $\{d_m\}_{m=1}^{|\mathcal{M}|}$ where each document $d_m$ is annotated with $T_m$ tags, i.e. $\mathcal{T}_m = \{t_{m,i}\}_{i=1}^{T_m}$, that belong to a tag vocabulary of size $T$. Furthermore, each document $d_m$ is a sequence of $N_m$ words, i.e. $(w_{m,1},..,w_{m,N_m})$ coming from a word vocabulary of size $V$. Hereafter, we drop the document index $m$ from its words and tags indices as it should be clear from the context. Given such data, our goal is to learn a $D$-dimensional continuous vector representations of words and tags that can be used to rank tags for a given document. These representations form matrices $\mathbf{U} \in \mathbb{R}^{\{D \times V\}}$ and $V \in \mathbb{R}^{\{D \times T\}}$ Our pro-posed relevance feedback based representation learning model defines each tag (i.e. its *embedding*) that appears in the context of document $m$ as:

$$\mathbf{V}_{(:,t)} = \sum_i^{N_m} \frac{f(\mathbf{U}_{(:,w_i)}, \mathbf{V}_{(:,t)})}{N_m} \mathbf{U}_{(:,w_i)} \qquad (1)$$

Here, $f(\mathbf{U}_{(:,w_i)}, \mathbf{V}_{(:,t)})$ (referred as $f(\mathbf{w,t})$ hereafter) defines a similarity (or relevance) function between $w$ and $t$ based upon the word and tag representations, respectively. Note that unlike topic models, $f(.)$ does not corresponds to a mixture model providing latent variable based association between a word and tag. Instead, the relevance model directly computes a semantic similarity between word and tag in representation space. Moreover, division by document length $N_m$ avoids learning bias towards longer document. The recursive representation of tags can be seen as a feedback loop (depicted in Figure 1), and is learned iteratively. Also, note that if we set $f(\mathbf{w,t}) = 1$, Eq. 1 refers to WSABIE [Weston *et al.*, 2011].

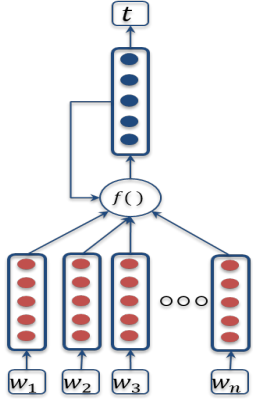

Figure 1: A recursive neural embedding model. Blue and red nodes denote tags and words representations resp. See text for $f(.,.)$

In this work, we define three similarity functions in the order of complexity and expressive power:

- *Scalar Product:* We define $f(\mathbf{w,t}) = g(\mathbf{w}^\top \mathbf{t})$, where $g = sigm$ as the scalar product of word and tag representations. This similarity measure weights each word based upon how similar it is with tag in the original representation space. The associated non-linearity helps *switch-on* relevant features while marginalizing non-relevant ones. This similarity model does not assume any complex interaction between word and tag.

- *Single Feedforward Layer:* We define

$$f(\mathbf{w,t}) = \mathbf{r}^\top g(\mathbf{R_1.w} - \mathbf{R_2.t})$$

where $g = sigm$ and $\mathbf{r} \in \mathbb{R}^{\{1 \times k\}}$, $\mathbf{R_1} \in \mathbb{R}^{\{k \times D\}}$, $\mathbf{R_2} \in \mathbb{R}^{\{k \times D\}}$ are the parameters of the feedforward network for similarity function. This similarity function computes similarity in a linearly transformed low dimensional space (similar to PCA).

- *Neural Tensor Layer:* We define

$$f(\mathbf{w,t}) = \mathbf{r}^\top g(\mathbf{w}^\top \mathbf{M_t}^{[1:k]} \mathbf{t} + \mathbf{b_t})$$

where $\mathbf{r} \in \mathbb{R}^{\{1 \times k\}}$, $\mathbf{M_t} \in \mathbb{R}^{\{D \times D \times k\}}$, $\mathbf{b_t} \in \mathbb{R}^{\{1 \times k\}}$ are the parameters of the tensor layer. Neural tensor layer computes the similarity between a tag-word pair in multiple *contexts* (or senses) where each slice of $\mathbf{M_t}$ corresponding to tag $t$ defines a context along which to compute the similarity score. For example, with context *technology*, the similarity of tag *apple* will be higher to word *device* as opposed to word *fruit*. Also, this tensor layer let word and tag interact multiplicatively along different contexts and produce context dependent similarity score. Fig. 2 depicts the neural tensor based similarity function.

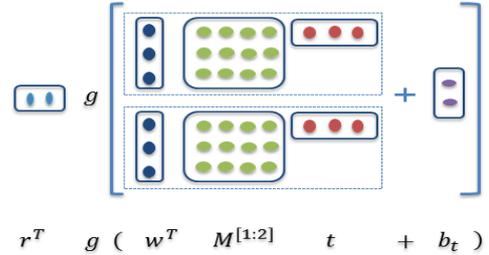

Figure 2: Visualization of the Neural Tensor Layer for similarity function $f(\mathbf{w,t})$

Although tensor factorization has been shown to be successful in modeling various ternary relations (e.g. entity relation modeling [Socher *et al.*, 2013], word sense disambiguation [Liu *et al.*, 2015]), introducing tensor based parameters per tag is computation prohibitive and, due to addition of millions of parameters, can lead to overfitting. Therefore, we make two modifications: (i) we use same tensor for all the tags, and (ii) apply a tensor factorization approach that factorizes each tensor slice as the product of two low-rank matrices. Formally, each tensor slice $M^{[i]} \in \mathbb{R}^{\{D \times D\}}$ is factorized into two low rank matrices, $P^{[i]} \in \mathbb{R}^{\{D \times p\}}$ and $Q^{[i]} \in \mathbb{R}^{\{p \times D\}}$, i.e.

$$M^{[i]} = P^{[i]} Q^{[i]}, 1 \leq i \leq k, p << D$$

**Objective function and Optimization**

Based upon the tag representation mentioned above, we define a scoring function for a given document tag pair, $(m,t)$ as

$$\mathcal{L}(m,t) = [f(\mathbf{U}, \mathbf{V}_{(:,t)})]^\top \mathbf{I_m} [\mathbf{U}^\top \mathbf{V}_{(:,t)}] \qquad (2)$$

where $\mathbf{I_m} \in \mathbb{R}^{\{V \times V\}}$ defines a diagonal matrix where diagonal entries are normalized word counts appearing in document $m$.

We use the contrastive max-margin criterion [Socher *et al.*, 2013] to train our model [1]. The main idea is that each pair $(m, t)$ coming from the training corpus should receive a higher score than a pair $(m, t^-)$ in which tag $t^-$ is a random tag. Let the set of all parameters be $\Psi$, we minimize the following objective:

$$\mathcal{J}(\Psi) = \sum_{\mathcal{M}} \sum_{\mathcal{M}^-} max\{0, 1 - \mathcal{L}(m, t) + \mathcal{L}(m, t^-)\} + \lambda ||\Psi||_2^2 \quad (3)$$

where $\mathcal{M}$ is the set of doc-tag pairs from training corpus (and $\mathcal{M}^-$ corrupted pairs, i.e. $(m, t^-)$'s where $t^-$ is picked randomly) and we score the correct pair higher than its corrupted one up to margin of 1. For each correct triplet we sample $neg = 5$ random corrupted triplets. We used standard $L2$ regularization of all the parameters, weighted by the hyperparameter $\lambda$.

## Experiments

**Datasets:** For our experiments, we use two real-world datasets with one from Citeulike [2] and one from MovieLens [3] [4]:

*CiteULike:* Our first dataset is originally from [Wang and Blei, 2011] and is collected from CiteULike database dump for over six years from 2004 to 2010. This dataset was originally designed for personalized document recommendation and consists of documents tagged by users with at least 10 articles. [Wang *et al.*, 2013] further extended this dataset with corresponding tag information from citeulike website. Each article is mapped to papers that are indexed in CiteSeerX to extract their titles and abstracts, resulting in 16980 articles, 7386 tags, and 204987 tag-item pairs. Tags with frequencies less than 5 have also been removed from the dataset.

*MovieLens dataset* : For our second dataset, we combined two publicly available datasets:(1) movie ratings dataset MovieLens 10M [5], consisting of movie ratings as well as tags for around 10K movies with around 25K unique tags; (2) movie synopses data set from Internet Movie DataBase (IMDB) [6]. After mapping movies from MovieLens data with IMDB data, there are 5,333 distinct movies plots and 15,558 distinct tags.

## Baselines

**Latent Space based Representation:** Latent space based tag recommendation methods typically combine Latent Dirichlet Allocation [Blei *et al.*, 2003] that learns item representations in one latent space with probabilistic matrix factorization (PMF) [Salakhutdinov and Mnih, 2008] that learns from tag-item co-occurrence matrix in another latent space. Collaborative topic regression (CTR) [Wang and Blei, 2011] learns from these two latent spaces simultaneously and achieve significant improvements over content, co-occurrence based, and hybrid methods for tag recommendation methods [Wang *et al.*, 2013; Wang and Blei, 2011; Purushotham *et al.*, 2012; Wang *et al.*, 2015]. Therefore, we use *CTR* as our topical representations based tag recommendation baseline.

**Neural network based Representations:** For our neural network based baselines, we use two broad categories of baselines:

*Generative Neural Architectures:* Generative neural architecture such as *Probabilistic Stacked Denoising Autoencoders* [Wang *et al.*, 2015] are generative variants of *denoising autoencoders* [Vincent *et al.*, 2010] that tries to reconstruct a noisy version of the input by learning to predict clean input with an intermediate low-dimensional bottleneck layer. [Wang *et al.*, 2015] extended the framework of denoising autoencoders by (i) proposing a deep architecture, and (2) assigning a data generating distributions to decoding layer of autoencoders and maximizing posterior probability of denoised input. [Wang *et al.*, 2015](referred as *SDAE*) shows significant improvements over generative neural architectures for tag recommendation. We use *SDAE* as our second baseline.

*Supervised Neural Architectures:* As mentioned earler, WSABIE [Weston *et al.*, 2011] is one of the most popular supervised embeddings based representation learning approach and it's variants has been shown to perform well on various annotation and recommendation task e.g. image annotation with labels [Weston *et al.*, 2013c], personalized tag recommendation [Denton *et al.*, 2015], etc. Therefore, we use following variants of WSABIE (along with itself) as our baselines:

- *TagSpace:* TagSpace [Weston *et al.*, 2014] extends WSABIE with a convolution neural network (CNN) over word sequences in a document to get a document representation. WSABIE does not learn from sequence information present in a document which is captures by a CNN with variable length window over document's text. *TagSpace* has shown significant improvements over WSABIE for large scale hashtag representation [Weston *et al.*, 2014].

- *Affinity Weighted Embeddings: AWE* [Weston *et al.*, 2013b] provides a framework similar to ours which is suitable for feature modalities other than text. Their approach differs from ours in two main ways: (1) AWE does not account for document length normalization which penalizes shorter documents. (2) AWE does not account for complex multiplicative interaction between tags and words due to document's context.

- *CSRW:* **C**ompositional **S**emantic **R**elevance **W**eighted embeddings are our proposed repre-

---
[1] Note that other choices of ranking criterion include WARP [Weston *et al.*, 2011], however, we select max-margin criteria for its wide applicability and fair to baselines.

[2] http://www.citeulike.org/faq/data.adp

[3] http://grouplens.org/datasets/movielens/

[4] Datasets used and code are available at https://github.com/ktsaurabh/recursive_WSABIE

[5] grouplens.org/datasets/movielens/

[6] availabe at: ftp.fu-berlin.de/pub/misc/movies/database/

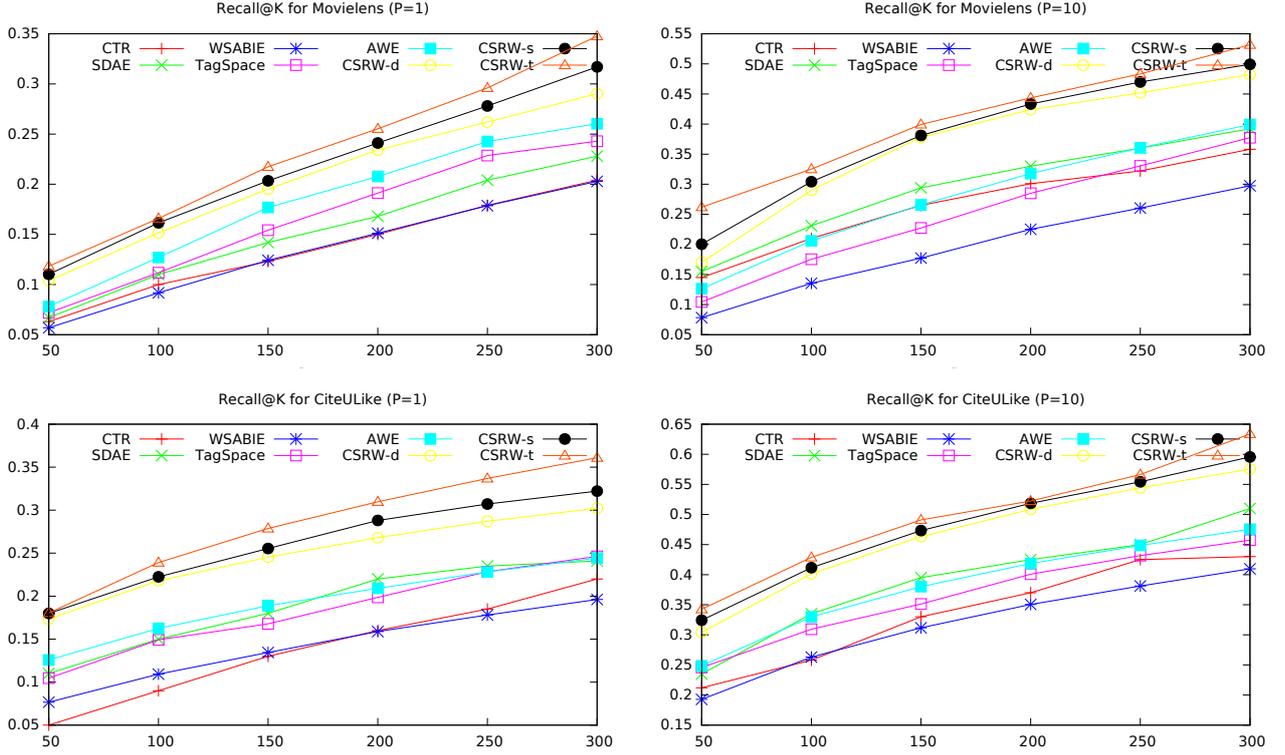

Figure 3: Performance comparison of all methods based on recall @ K for (i) Movielens dataset (top) and (ii) CiteULike dataset (bottom) with different sparsity ($P$) settings.

|  |  | **CTR** | **SDAE** | **WSABIE** | **TagSpace** | **AWE** | **CSRW-d** | **CSRW-s** | **CSRW-t** |
|---|---|---|---|---|---|---|---|---|---|
| **Movielens** | **P=1** | 1.01% | 1.23% | 1.19% | 1.26% | 1.34% | 1.52% | 1.62% | **1.83%** |
|  | **P=10** | 1.21% | 1.37% | 1.32% | 1.52% | 1.63% | 1.94% | 2.53% | **3.70%** |
| **CiteULike** | **P=1** | 1.41% | 1.98% | 1.61% | 2.14% | 2.44% | 3.38% | 3.41% | **3.56%** |
|  | **P=10** | 2.64% | 3.65% | 2.85% | 3.33% | 3.97% | 5.26% | 5.81% | **6.23%** |

Table 1: Mean Average Precision (MAP) results for two datasets. $P$ corresponds to sparsity setting.

sentations for tag recommendation. We use three variants of CSRW corresponding to three relevance functions defined in previous section. That is, (i) *CSRW-d* for dot product based similarity function, (ii) *CSRW-s* for single layer feedforward network based similarity function, and (iii) *CSRW-t* for neural tensor based similarity function.

**Evaluation Settings**

In each dataset, similar to [Wang *et al.*, 2013; 2015], $P$ items associated with each tag are randomly selected to form the training set and all the rest of the dataset is used as the test set. P is set to 1 and 10, respectively, to evaluate and compare the models under both sparse and dense settings in the experiments. For each value of $P$, the evaluation is repeated five times with different randomly selected training sets and the average performance is reported. Following [Wang *et al.*, 2015; Wang and Blei, 2011; Wang *et al.*, 2013], we use recall as the performance measure since the rating information appears in the form of implicit feedback [Rendle *et al.*, 2009], which means a zero entry may be due to irrelevance between the tag and the item or the user's ignorance of the tags when tagging items. As such, precision is not suitable as a performance measure. Like most recommender systems, we sort the predicted ratings of the candidate tags and recommend the top K tags to the target item. The recall @ K for each item is defined as:

$$\text{Recall @ K} = \frac{\text{number of tags the item is associated with in top K}}{\text{total number of tags the item is associated with}}$$

For evaluating overall tag recommendation performance, we also use Mean Average Precision (MAP) as another metric. Apart from taking overall recall into consideration, it also measures overall precision of recommendation system. MAP is defined as:

$$\text{MAP} = \frac{1}{|\mathcal{M}|} \sum_{m \in \mathcal{M}} \frac{1}{|t_m|} \sum_{j \in t_m} \text{Precision}(R_{mj})$$

| Keyword | Most Similar Tags |
|---|---|
| *language* | knowledge, strategies, culture, behavioral, decision_making, identity, psychology, personality, response |
| | syntax, grammar, linguistics, semantics, teaching, context, dsl, read, operational, haskell, language_acquisition |
| *neural* | perception, visual, function, coding, spiking, neural_coding, attention, learning, vision, neuroscience, interaction, memory, cortex, retina |
| | neural, risk, computational, missing, time_series, discovery, large, communities, analysis, correlation, practice, missing_data |
| *machine learning* | statistics, stats, algorithm, classification, machinelearning, bayesian, model, learning, regression, inference, modelling, large_scale |
| | genome, disease, genomics, neuroscience, memory, association, methods, genome_analysis, sequence, human_diseases, genome_sequencing, human_genome |
| *markov* | sampling, markov, statistics, mcmc, monte_carlo, random, distribution, chain, testing, matrix, likelihood, free_energy, hmm, stochastic |
| | bayesian_network, structure, bayesian, variable, cooperative, bayes, structural, dynamic, graphical_models, feature, dynamic_programming, inference |

Table 2: Nearest tag neighbors of certain keywords. Duplicate variants in plural forms are removed. Similarity is defined based upon a certain slice of tensor as described in text.

where $t_m$ is the set of all the test tags for document $m$ and $R$ defines a ranking order on all the test tags of a document.

**Parameter settings:**

For CTR and SDAE, we set parameters as described by [Wang *et al.*, 2013] and [Wang *et al.*, 2015] respectively. Specifically, for CTR we fix the topic size to be 50 and choose hyperparameter based upon a validation set. For SDAE, we use a 2-layer architecture (described as setting '20000-200-50-200-20000' in [Wang *et al.*, 2015]). For WSABIE and AWE, we use the same settings as for CSRW based approaches. That is, we find the best performing dimension size (=200) based upon a grid search with a 5-fold cross-validation scheme. We use same objective function (i.e., Eq.2) for all WSABIE based baselines and CSRW models. We use SGD as our learning algorithm. For CSRW-s, we use the layer size, i.e., k = 16. For CSRW-t, we use tensor slices size, i.e., k=4 and factorization size, i.e., p = 16.

**Results:** Fig. 3 shows the recall @ K, where K ranges from 50 to 300, for two datasets. For recall, our 95% confidence interval indicates between 3-5 %, for $P = 10$, and 4-6%, for $P = 1$, deviation from mean reported (higher for Movielens data), indicating statistical significance of results. Evidently, CSRW based approaches outperforms WSABIE variants as well as topic modeling based CTR and probabilistic deep network based SDAE. Interestingly, performance of SDAE is comparable to variants of WSABIE (i.e., AWE and TagSpace). AWE outperforms TagSpace indicating that non-linear weighting of word features provides a better fit to data. Furthermore, the difference between AWE and CSRW-d (which is equivalent to document normalized AWE) clearly highlights the significance of document normalization. The good performance of neural tensor based relevance layer can be attributed to a better capacity for tags and words to interact multiplicatively. Sparsity has effect on the performance of all the models as expected. However, relative improvements for CSRW models are consistent with improvements upto 36% for higher K and 47% for lower K's for $P = 1$ and upto 32% for higher K and 65% for lower K's for $P = 10$. Table 1 shows the MAP for two datasets which follows a similar trend. Since MAP accommodates for Precision as well as overall performance of recommender system, CSRW provides significant improvements overall.

To elaborate the multi-factor interaction of tags and words, we present most similar tags to words based upon different slices of tensor $\mathbf{M}^{[\mathbf{k}]}$. Table 2 shows anecdotal evidences of separation of senses of tags that corresponds to same keywords. Here, we rank tags based upon $f(\mathbf{w}, \mathbf{t})[i] = \mathbf{w}^\top \mathbf{P}^{[i]} \mathbf{Q}^{[i]\top} \mathbf{t}$ and show top ranked tags in representative $i$. Clearly, the tags tend to cluster into type of usages, for example, tags corresponding to the keyword "neural" are clustered into *computation* and *biological* neural networks.

## Conclusion

We have presented a novel neural model based architecture that takes advantage of tags associated with documents to find meaningful representation of content and tags. We have presented a relevance feedback based neural representation learning framework where tags representations are learned by weighting words representations with its associated tags in a recursive fashion. We have shown that our modeling scheme outperform several state of the art baselines for tag recommendation task.